\documentclass[aip,apl,showpacs,preprint,amsmath,amssymb,superscriptaddress]{revtex4-1}

\usepackage{graphicx}
\usepackage{dcolumn}
\usepackage{amsmath}
\usepackage{bm}
\usepackage{array}
\usepackage{natbib}
\usepackage{epstopdf}
\usepackage{color}

\begin{document}

\title{Probing limits of STM field emission patterned Si:P $\delta$-doped devices}

\author{M.~Rudolph}
\affiliation{Sandia National Laboratories, Albuquerque, NM, 87185, USA}

\author{S.~M.~Carr}
\affiliation{Sandia National Laboratories, Albuquerque, NM, 87185, USA}

\author{G.~Subramania}
\affiliation{Sandia National Laboratories, Albuquerque, NM, 87185, USA}

\author{G.~Ten~Eyck}
\affiliation{Sandia National Laboratories, Albuquerque, NM, 87185, USA}

\author{J.~Dominguez}
\affiliation{Sandia National Laboratories, Albuquerque, NM, 87185, USA} 

\author{T.~Pluym}
\affiliation{Sandia National Laboratories, Albuquerque, NM, 87185, USA}

\author{M.~P.~Lilly}
\affiliation{Sandia National Laboratories, Albuquerque, NM, 87185, USA}
\affiliation{Center for Integrated Nanotechnologies, Sandia National Laboratories, Albuquerque, NM, 87185, USA}

\author{M.~S.~Carroll}
\affiliation{Sandia National Laboratories, Albuquerque, NM, 87185, USA}

\author{E.~Bussmann}
\affiliation{Sandia National Laboratories, Albuquerque, NM, 87185, USA}

\date{\today}

\begin{abstract}
Recently, a single atom transistor was deterministically fabricated using phosphorus in Si by H-desorption lithography with a scanning tunneling microscope (STM).  This milestone in precision, achieved by operating the STM in the conventional tunneling mode, typically utilizes very slow ($\sim\!10^2~\mathrm{nm^2/s}$) patterning speeds.  By contrast, using the STM in a high voltage ($>10~\mathrm{V}$) field emission mode, patterning speeds can be increased by orders of magnitude to $\gtrsim\!10^4~\mathrm{nm^2/s}$.  We show that the rapid patterning negligibly affects the functionality of relatively large micron-sized features, which act as contacting pads on these devices.  For nanoscale structures, we show that the resulting transport is consistent with the donor incorporation chemistry enhancing the device definition to a scale of $10~\mathrm{nm}$ even though the pattering spot size is $40~\mathrm{nm}$.
\end{abstract}

\maketitle

%\section{Introduction}

As scaling limitations become more problematic in classical CMOS technologies, new fabrication techniques are being examined both to assist and extend Moore's law, as well as to explore beyond Moore's law computation schemes such as quantum computing.\cite{Kane1998}  Recently, atomic precision fabrication of a single atom transistor was reported using H-desorption lithography patterned with a scanning tunneling microscope (STM).\cite{Fuechsle2012}  The demonstration of robust few-donor and single-donor devices using this technique opens the door to many near-atomic-precision transistor and quantum bit (qubit) designs in Si or Ge.\cite{Fuechsle2010, Mahapatra2011, Buch2013, Weber2014, Watson2014, Scappucci2011}

\begin{figure}[]
\includegraphics[width=85mm]{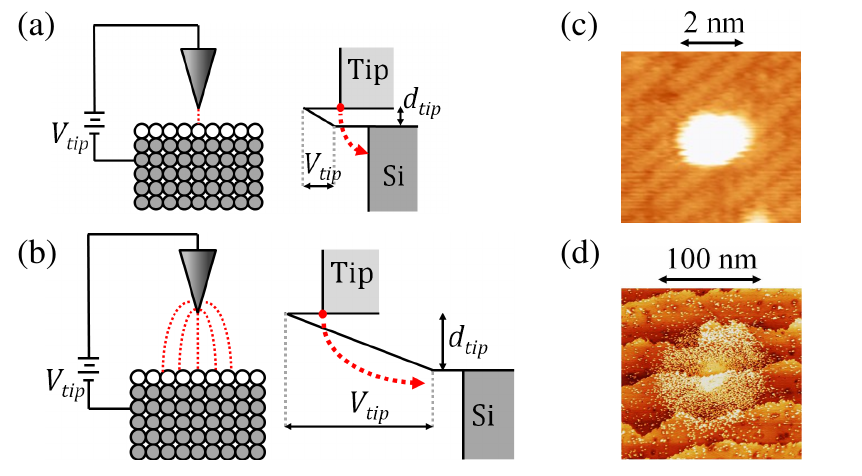}
\caption{Schematic energy diagrams of the STM operating in a tunneling mode (a) and a HVFE mode (b).  Characteristic lithographic spots for the tunneling (c) and HVFE (d) modes.}
\label{fig:stmmode}
\end{figure}

This paper demonstrates that high voltage field emission (HVFE) lithography is a way to significantly decrease lithographic write times while keeping the integrity of the donor devices.  Historically, the STM patterning step is typically performed at relatively low tip-substrate voltages $V_{tip}$ ($3-5~\mathrm{V}$), which we will call the tunneling mode.\cite{Fuechsle2010, Mahapatra2011, Fuechsle2012, Buch2013, Lyding1994, Shen1995, Adams1996, Ballard2013} In this mode, the tip-substrate distance $d_{tip}\approx 1~\mathrm{nm}$, and tip-substrate bias $V_{tip}$ is comparable to the work function of the tip (Fig~\ref{fig:stmmode}(a)), resulting in an $\mathrm{\AA}$ngstrom-sized lithographic spot (Fig~\ref{fig:stmmode}(c)) ideal for atomic-precision-patterning.  Removing a H-atom requires multi-electron capture for $V_{tip}<8V$,  and the interaction cross-section to desorb a H-atom decreases exponentially with decreasing $V_{tip}$ in the tunneling mode, making this an inefficient process.  Additionally, the writing current is limited to a few nanoamps to prevent tip instability. With these two limitations, the maximum possible writing speed is limited to the order of $1~\mathrm{\mu m/s}$.  In practice much lower speeds ($\sim 100$ nm$/$s) are imposed by the bandwidth ($\sim\!100~\mathrm{Hz}$) of the STM feedback loop, limiting the areal patterning rate to $\sim\!10^2~\mathrm{nm^2/s}$.  To overcome the speed limitation, previous works have utilized a low-voltage field emission (LVFE) mode, where $V_{tip}\approx 5-10~\mathrm{V}$, for which the areal patterning rate increases by less than an order of magnitude. \cite{Shen1995, Adams1996, Ballard2013, Buch2013}  Here the interaction cross-section approaches its saturation point, where the H-desorption becomes a single electron process.

Adams {\it{et al.}} recognized the potential to achieve even more rapid H-desorption in a high voltage field emission (HVFE) mode, but this technique has yet to be extended to donor device fabrication.\cite{Adams1996}  In this mode, $d_{tip}>3~\mathrm{nm}$ and $V_{tip}>10~\mathrm{V}$ such that the work function of the tip is insignificant compared to $V_{tip}$ (Fig.~\ref{fig:stmmode}(b)).  The maximum electric field between the tip $(V_{tip}/d_{tip})$ is comparable to the tunneling mode, but the radial decay of the field magnitude is proportional to $d_{tip}$ for constant $V_{tip}/d_{tip}$, thereby enlarging and diffusing the patterning spot (Fig~\ref{fig:stmmode}(d)).  In addition, the large $d_{tip}$ allows the STM feedback loop to be removed and extends the areal pattern speed to $>10^4~\mathrm{nm^2/s}$, a more than two order of magnitude speed up over patterning in the tunneling mode.

In many donor device layouts, atomic precision is only necessary for a small active region of the device (micron sized donor pads are still required for robust electrical connections by vias and interconnects), so utilizing the large HVFE mode spot size can dramatically improve patterning speeds.  In this paper, we show that using the STM in a HVFE mode allows for patterning large features at speeds orders of magnitude faster than possible with the STM operating in the standard tunneling or LVFE mode while negligibly effecting the $\delta$-doped donor layer.  We then explore the donor device feature size limitations using HVFE mode patterning by transport measurements of a 22 nm tunnel gap device.  We show that the donor incorporation chemistry leads to similar device characteristics to atomically-precise tunnel mode devices\cite{Pok2011} despite the HVFE patterning spot being diffuse.

\begin{figure}[]
\includegraphics[width=85mm]{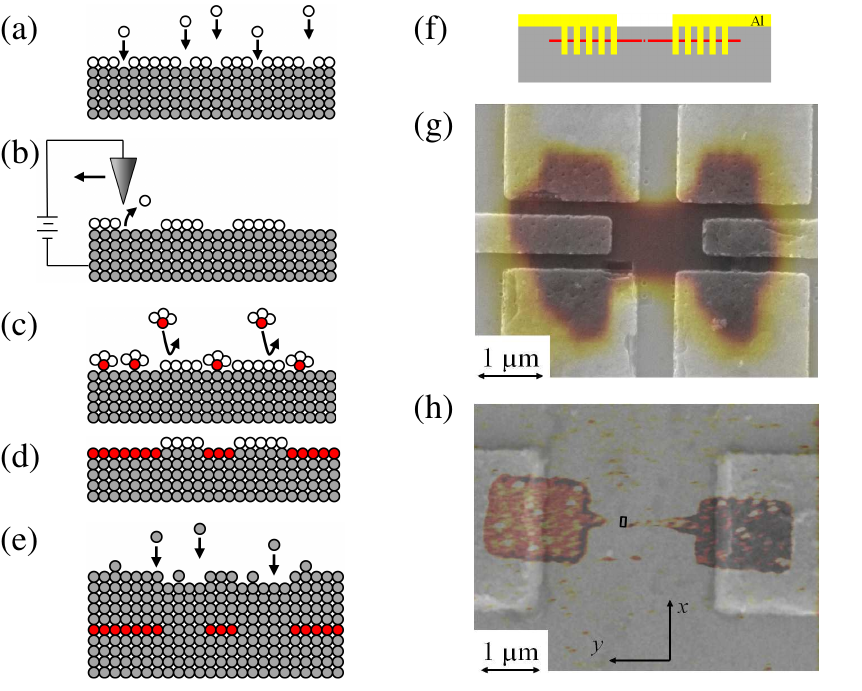}
\caption{(a-f) The notional fabrication process flow.  Scanning capacitance images of the buried donor Hall bar (g) and tunnel junction (h) devices overlain on scanning electron images of the Al leads.}
\label{fig:processflow}
\end{figure}

%\section{Fabrication}

The devices in this study were fabricated in a custom UHV ($5 \times 10^{-10}~\mathrm{Torr}$) STM system equipped with atomic-H and thermal Si sources.  Si(001) substrates are loaded into the STM and flash annealed at $\sim\!1250~\mathrm{^{\circ}C}$ for 15 seconds several times to prepare atomically clean and flat Si(001)($2\times 1$) surfaces.  To prepare a monohydride surface resist, the sample is held at $\sim\!350~\mathrm{^{\circ}C}$ near a $2000~\mathrm{^{\circ}C}$ W filament in the presence of $10^{-6}~\mathrm{Torr}$ of $\mathrm{H_2}$ (Fig.~\ref{fig:processflow}(a)).  The device mask is defined by H-lithography, where adsorbed H atoms are selectively desorbed by electrons tunneling to the biased substrate from the grounded STM tip (Fig.~\ref{fig:processflow}(b)).  Devices are then doped by a subsequent exposure to $2 \times 10^{-8}~\mathrm{Torr}$ of $\mathrm{PH_3}$ for 5 minutes during which $\mathrm{PH_3}$ molecules adsorb to open Si sites while H-depassivated Si sites remain inert ((Fig.~\ref{fig:processflow}(c)).  A 30 second anneal at $\sim\!400~\mathrm{^{\circ}C}$ incorporates the P into the outermost layer of the Si lattice ((Fig.~\ref{fig:processflow}(d)).  The H-mask remains intact during the anneal, blocking lateral diffusion of the P atoms across the surface and providing atomically precise donor doping in all three dimensions.\cite{Flowers1998, Bowler2000}  Finally, a $30~\mathrm{nm}$ thick epitaxial Si capping layer is grown at $250~\mathrm{^{\circ}C}$ to bury the patterned P $\delta$-doped layer, which protects the device from surface effects ((Fig.~\ref{fig:processflow}(e)).  Electrical connections to the buried $\delta$-layer device are achieved by conventional electron beam lithography, with $100~\mathrm{nm}$ deep etched via holes filled with $100~\mathrm{nm}$ thick patterned Al ((Fig.~\ref{fig:processflow}(f)).  Composite images of the  completed devices for this study are displayed in Figs.~\ref{fig:processflow}(g) and \ref{fig:processflow}(h), highlighting the necessary $\sim\!1~\mathrm{\mu m^2}$ overlap between the $\delta$-doped layer and the top-side metallization for ohmic contact.  The buried structures are evident in scanning capacitance measurements due to the $\sim\!6$ orders of magnitude larger donor concentration in the device than in the substrate.\cite{Bussmann2014}

We patterned two devices with the accelerated writing speed possible from the HVFE mode.  The tip current was $1~\mathrm{nA}$ for both devices.  The first is a micron-sized Hall bar to examine the activation, mobility, and contact resistance in $\delta$-doped layers patterned in the HVFE mode (Fig.~\ref{fig:processflow}(g)).  The second is a $22~\mathrm{nm}$ long tunnel gap between two leads with widths ~$\lesssim 40~\mathrm{nm}$, shown in Fig.~\ref{fig:processflow}(h), where the active region is inside the rectangle.  Here we study the scaling limitations of using the larger patterning spot size inherent in the HVFE mode to define the donor device.  This includes identifying the sharpness of the H-mask edges and understanding the impact of the increase in random sites being depassivated near the target exposed region on the donor placement.  Electrical characterization of the devices are performed at $\sim\!4~\mathrm{K}$.  At this temperature, the carriers in the low-doped substrate are frozen out, and conduction only occurs through the patterned $\delta$-doped region.

%\section{Experiments}

The HVFE patterned Hall bar is $\sim\!2~\mathrm{\mu m}$ long and $1~\mathrm{\mu m}$ wide, with six $1\times 1.5~\mathrm{\mu m^2}$ contacting pads extending from the active region, shown in Fig.~\ref{fig:processflow}(g).  During STM lithography, $d_{tip}\approx 10~\mathrm{nm}$ and $V_{tip}=110~\mathrm{V}$.  The patterning spot size was about $50~\mathrm{nm}$ under these conditions.  The total STM write time for the doped region ($\sim\!10~\mathrm{\mu m^2}$) is less than 10 minutes, whereas it would have taken 10s of hours if patterned at typical speed in tunneling mode.  A total of six ohmic contacts were made to the buried device ($R_{contact} \approx 8~\mathrm{k\Omega}$) to measure both the resistivity and Hall signal.  Hall and van der Pauw measurements indicate the electron density in the $\delta$-doped layer to be $n = 7\pm 1 \times 10^{13}~\mathrm{cm^{-2}}$ with a mobility of $\mu = 50\pm 20~\mathrm{cm^2/Vs}$.  The electron density is large enough to provide ohmic contacts to the $\delta$-doped layer, which is the sole purpose of the large features in these $\delta$-doped devices.  Higher electron densities of $2\times 10^{14}~\mathrm{cm^{-2}}$ have been achieved when blanket doping Si(001) with a single $\mathrm{PH_3}$ dose.\cite{Goh2006, Mckibbin2014}  Reasons for the reduced density in our devices may be the measurement uncertainty from the non-ideal Hall bar geometry, incomplete desorption of the H-mask using the HVFE mode with an unoptimized dwell time,  possible surface repassivation due to the speed of the HVFE patterning, and dopant diffusion due to uncertainty in the incorporation temperature.  The electron density is, nevertheless, well above the metal-insulator transition and therefore provides an ohmic region to which contacts can be made, demonstrating that micron-sized device features can be produced using the HVFE mode.

\begin{figure}[]
\includegraphics[width=85mm]{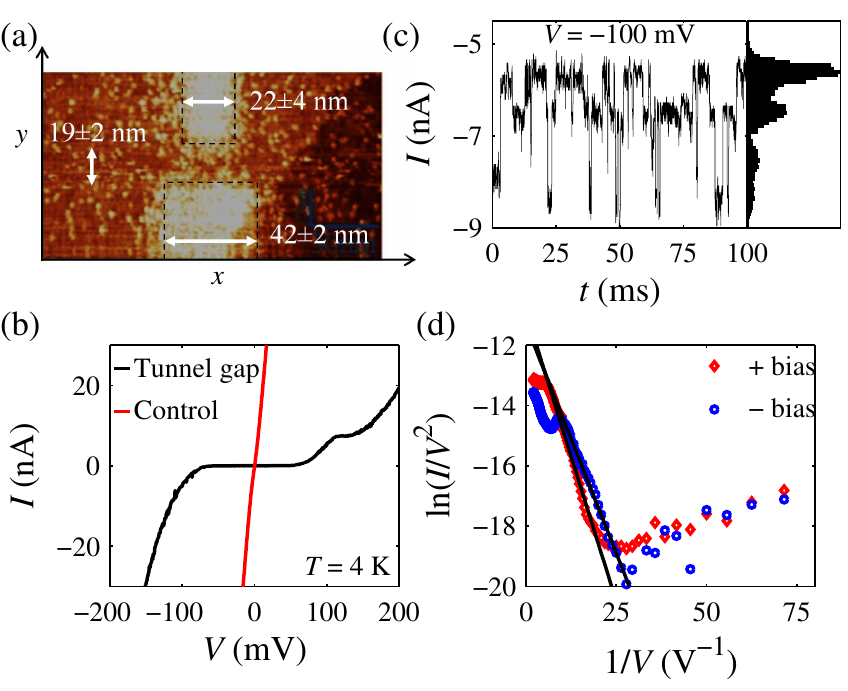}
\caption{(a) STM image of the H-desorbed mask of the tunnel gap before the P incorporation.  (b) Two-terminal measurement of the current dependence on voltage of the tunnel gap device and a control case testing the contacts. (c) Random telegraph signal when the tunnel gap is biased to $-100~\mathrm{mV}$, with the adjacent histogram indicating 4 dominant metastable levels. (d) Fowler-Nordheim plot of (b).}
\label{fig:tunneldata}
\end{figure}

To probe the lower bound on the device feature sizes that can be patterned with the large diffuse spot in the HVFE mode, and to assess the effects of adventitious partial H-desorption at the periphery of these structures, we have fabricated a nanoscale tunnel junction device.  A tunnel barrier's $I$-$V$ characteristics are very sensitive to the gap size and can be calibrated to previous measurements of tunnel barriers made with atomic precision fabrication.  Using the HVFE mode, two leads were patterned with a separation of $\sim\!20~\mathrm{nm}$.  The resulting structure is imaged after the HVFE patterning step, and locations where the H-mask is removed appear as bright regions in the STM image (Fig.~\ref{fig:tunneldata}(a)).  The narrower top lead was written by a single pass of the STM tip, while the wider bottom lead was written by two passes.  The edge of the H-mask is speckled due to the broad diffuse electron beam in the HVFE mode.

The incorporation of P donors through adsorption of $\mathrm{PH_3}$ requires at least three sequential depassivated dimers along a dimer row.\cite{Wilson2004, Warschkow2005}  There is an observed decaying efficiency of H-desorption away from the tip's center (i.e., the location of highest current from the tip), which produces many regions that have less than three neighboring dimers that are stripped of H.  The electrical dimensions of the tunnel junction will, therefore, differ from the H-mask dimensions defined by any depassivated region because P donors cannot incorporate into regions with a low density desorption of H.  A central point of the work is to clarify how much contribution the spurious depassivation makes on the tunnel barrier conductance and whether an edge of the leads can be defined.

The $I$-$V$ characteristics of the tunnel gap are presented in Fig.~\ref{fig:tunneldata}(b).  A blockaded region, indicative of a  robust barrier, is observed for $|V| \lesssim 50~\mathrm{mV}$.  This contrasts the ohmic nature of the control, which measures the resistance of the contacts to the $\delta$-layer.  A second plateau is present for $V \approx 100~\mathrm{mV}$, which may indicate the onset of a secondary conduction path, most likely through a spurious donor island in the tunnel gap.  Figure~\ref{fig:tunneldata}(c) displays a time sweep of the device conduction for a bias of $-100~\mathrm{mV}$ exhibiting a random telegraph signal (RTS) with 4 visible metastable states, another indication of few donor islands present in the tunnel gap.  The metastable levels are highlighted in the histogram in Fig.~\ref{fig:tunneldata}(c).  The tunnel times for the transitions between these states are between 1 and $5~\mathrm{ms}$.

In order to analyze the effective tunnel barrier height in the gap, we present the $I$-$V$ data on a Fowler-Nordheim plot, $1/V~\mathrm{vs.}~\ln(I/V^2)$ in Fig.~\ref{fig:tunneldata}(d).  The minimum in the graph indicates the transition between trapezoidal barrier tunneling (large $1/V$) and triangular barrier tunneling (small $1/V$), giving a direct approximation of the barrier height.  Data from both positive and negative bias display a minimum at $1/V \approx 28~\mathrm{V^{-1}}$, corresponding to a barrier height of $V_b \approx 35~\mathrm{mV}$.  We also extract the barrier height from the slope of the triangular barrier tunneling region, which is $-4d\sqrt{2m^*V_b^3}/3\hbar q$, where $d$ is the tunnel gap distance and $m^*$ is the effective mass within the tunnel gap \cite{Jensen2003}.  We fit our data using $d=15~\mathrm{nm}$ and $m^*=0.19m_e$ (dominant Si effect mass when confined to 2-dimensions) and find $V_b=36~(42)~\mathrm{mV}$ for positive (negative) bias.  Here $d$ is taken to be the physical lead separation (discussed below) less twice the Bohr radius of the donors to provide the electrical separation of the leads.  The values of $V_b\approx40~\mathrm{mV}$ from both methods are consistent with one another.  It is also within about factor 2 from the expected barrier height of $70-100~\mathrm{mV}$,\cite{Qian2005, Carter2009, Fuechsle2012} which is reasonable since the 1-dimensional Fowler-Nordheim square barrier model only approximates our tunnel gap.  In reality, the barrier has a parabolic component, is 2-dimensional (the barrier width and length are comparable), and contains spurious donor potentials.  All of these considerations would increase the calculated $V_b$ and offer even better agreement with the expected $V_b$.

%\section{Discussion}

The STM image of the H-mask in Fig.~\ref{fig:tunneldata}(a) shows many open Si sites, however, transport data shows at most one additional direct resonance (i.e., a single additional island in series) and on the order of four parallel islands that contribute RTS under certain bias conditions.  This presumably arises from the donor incorporation chemistry, which requires at least three neighboring depassivated dimers to incorporate a single P atom, and thus Fig.~\ref{fig:tunneldata}(a) is not an accurate description of the device's electrical dimensions.  To model the device's electrical dimensions, we first analyze the scatter of the H-desorbed site (dangling bonds) in Fig.~\ref{fig:tunneldata}(a) and then map the donor incorporation sites.

\begin{figure}[]
\includegraphics[width=85mm]{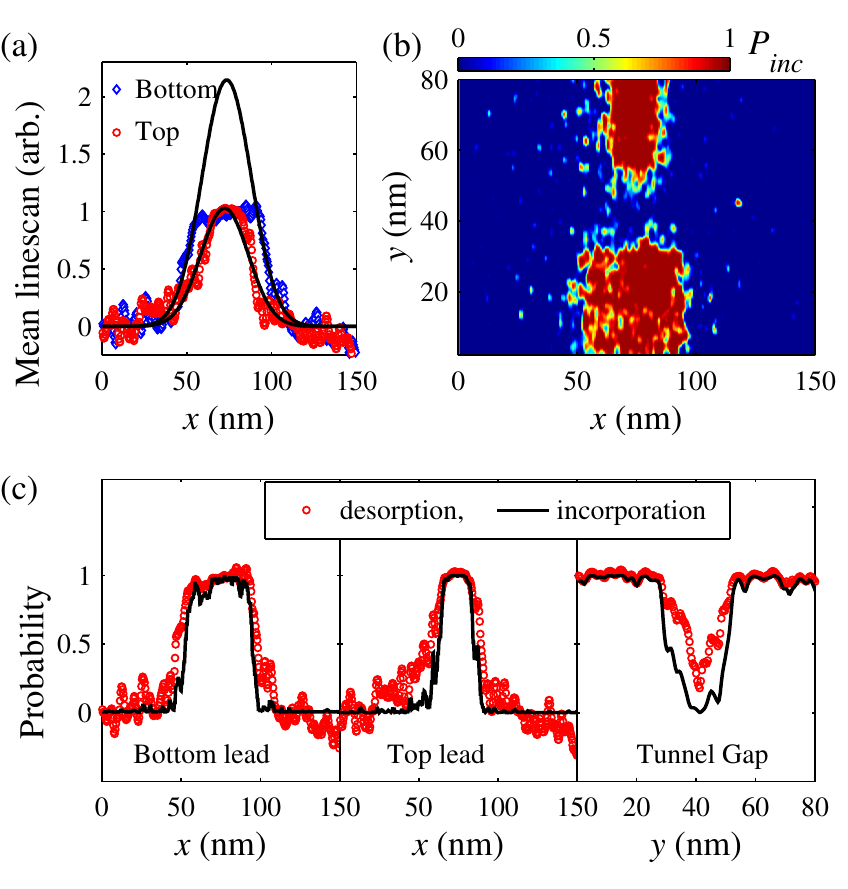}
\caption{(a) Normalized average of all STM linescans across the top and bottom leads with corresponding Gaussian fits to the edges.  (b) Contour map of the calculated probability of donor incorporation. (c) Average calculated device dimensions of the bottom lead, top lead, and tunnel gap.}
\label{fig:incorp}
\end{figure}

Figure~\ref{fig:incorp}(a) displays the average linescan of the top and bottom leads, highlighting the H-desorption tails that extend beyond the center of the leads.  The data has been normalized to pin the fully desorbed regions to 1 and the fully masked regions to 0.  We fit the edge profile of the average linescans to Gaussians, which we assume to be the shape of the HVFE spot, of the form $ae^{-(x-x_0)^2/r^2}$, where $r$, $a$, and $x_0$ are free parameters and $r$ denotes the spot radius.  The plateau regions where the H-desorption has saturated are not included in the fit.  We find $r$ to be $20.0~(20.6)~\mathrm{nm}$ for the bottom (top) lead.  We define the spot size as $2r\approx40~\mathrm{nm}$.  The amplitude of the Gaussian for the bottom lead is twice that of the top lead.  The top lead was written by one pass while the bottom lead witnessed two passes of the STM tip.  The averaged linescans agree with a Gaussian distribution.  The probability of H-desorption in this work, therefore, follows a normal distribution with radial distance.

The chemical pathway for a P donor to incorporate into the Si(001) surface lattice requires that at least three adjacent Si dimers be available on which the $\mathrm{PH_3}$ can adsorb, with the incorporation probability increasing with additional available dimers.\cite{Wilson2004, Warschkow2005}  We model the P incorporation by considering the individual incorporation probabilities for 3, 4, 5, and 6 dimer processes ($P_i$ with $3\le i\le 6$).\cite{Fuechsle2011}  The cutoff at 6 dimers is imposed since 6 available dimers already yield a 100\% incorporation probability.  These are then combined to calculate the P incorporation probability by
\begin{equation}
P_{inc} = P_6+(1-P_6)\{P_5+(1-P_5)[P_4+(1-P_4)P_3]\},
\label{eq:inc}
\end{equation}
 where each $P_i$ is calculated by using a grid reflecting the physical size of $2\times i$ Si atoms.  Figure~\ref{fig:incorp}(b) displays a contour plot of $P_{inc}$, where the normalized STM image from Fig.~\ref{fig:tunneldata}(a) is used as a measure of the probability of H-desorption for each Si atom.  The nominal device dimensions of the tunnel gap and leads are obtained from the average linecuts across the features, shown in Fig.~\ref{fig:incorp}(c).  We find the tunnel gap to be $19 \pm 2~\mathrm{nm}$ long and the top (bottom) lead to be $22 \pm 4~(42 \pm 2)~\mathrm{nm}$ wide.  The uncertainty is taken as the distance between $P_{inc}=0.9$ and $P_{inc}=1/e$.  While the STM spot for H-desorption is Gaussian in shape with a width of $40~\mathrm{nm}$, the donor incorporation edge falls off much more rapidly.

Figure~\ref{fig:incorp}(b) shows quite a few isolated islands of high $P_{inc}$, particularly near the immediate edge of the leads.  However, the Bohr radius of P donors in Si is $a_B \approx 2.5~\mathrm{nm}$, so any two donor regions within $5~\mathrm{nm}$ from each other are likely electrically connected.  Taking the Bohr radius into account, there are $\sim\!5$ electrically isolated donor islands in or near the tunnel gap.  The number of donors in the islands can be estimated by considering the size of the high-probability regions, which suggest islands with $>5$ donors are unlikely.  Further indication of the island sizes come from the tunnel times between the metastable states in the RTS.  Our observation of tunnel times between 1 and 5 ms is comparable to a 5 ms tunnel time measured in a similar device, where an intentional 4 P donor island was positioned 15 nm away from a single electron transistor.\cite{Buch2013}  Both the number of islands visible in Fig.~\ref{fig:incorp}(b) and the estimated island size are consistent with the 4 metastable levels with 1-$5~\mathrm{ms}$ tunneling times observed in the random telegraph signal in Fig.~\ref{fig:tunneldata}(c).

%\section{Conclusions}

In summary, we show STM H-desorption lithography in a high voltage ($>10~\mathrm{V}$) field emission mode for patterning large scale features for $\delta$-doped P donor devices.  This mode allows a speed-up in patterning time of a few orders of magnitude compared to previously reported methods for patterning donor devices.  This speed-up is due the high voltage providing a larger patterning spot size, a faster available scan rate, and a higher interaction cross-section for H-desorption.  We also show that a $22~\mathrm{nm}$ long tunnel barrier patterned with a $40~\mathrm{nm}$ wide spot size exhibits a relatively clean tunnel barrier despite $\sim\!50\%$ of the H in the tunnel barrier being adventitiously desorbed.  We model the P donor incorporation in this device and find that the electrical dimensions vary by a few nanometers even though the patterning spot has a Gaussian profile with a $20~\mathrm{nm}$ radius.  Features $\gtrsim 10~\mathrm{nm}$ can be rapidly patterned in the HVFE mode, greatly increasing the throughput of STM patterned Si:P $\delta$-doped devices.

This work was performed, in part, at the Center for Integrated Nanotechnologies, a U.S. DOE, Office of Basic Energy Sciences user facility. The work was supported by the Sandia National Laboratories Directed Research and Development Program. Sandia National Laboratories is a multi-program laboratory operated by Sandia Corporation, a Lockheed-Martin Company, for the U. S. Department of Energy under Contract No. DE-AC04-94AL85000.

%\bibliographystyle{apsrev}
%\bibliography{masterbib}

\end{document}